\documentstyle[12pt,epsfig]{article}
\def\beq{\begin{equation}}
\def\eeq{\end{equation}}
\def\bea{\begin{eqnarray}}
\def\eea{\end{eqnarray}}
\def\bq{\begin{quote}}
\def\eq{\end{quote}}

\def\gappeq{\mathrel{\rlap {\raise.5ex\hbox{$>$}}
{\lower.5ex\hbox{$\sim$}}}}

\def\lappeq{\mathrel{\rlap{\raise.5ex\hbox{$<$}}
{\lower.5ex\hbox{$\sim$}}}}

\begin{document}
\setcounter{page}{0}
\pagestyle{empty}
\begin{flushright}
{CERN-TH/2002-111}\\
{IPPP/02/28}\\
{DCPT/02/56}\\
{UCRHEP-T340}\\
{June 2002}\\
{Revised August 2002}\\
\end{flushright}
\vspace*{5mm}
\begin{center}
{\bf THE $\tau$-POLARISATION TEST FOR THE $H/A \rightarrow \tau^+\tau^-$ SIGNAL AT THE LHC} \\
\vspace*{1.5cm}
{\large S. Moretti$^{a,}$\footnote{stefano.moretti@cern.ch}
and D.P. Roy$^{b,}$\footnote{dproy@theory.tifr.res.in}}\\[4mm]
{\em $^{a}$ CERN Theory Division, CH-1211 Geneva 23, Switzerland\\
{\rm and}\\
Institute for Particle Physics Phenomenology,
University of Durham, Durham DH1 3LE, UK}\\[4mm]
{\em $^{b}$ Tata Institute of Fundamental Research, Mumbai, 400 005, India\\
{\rm and}\\
Physics Department, University of California, 
Riverside, CA 92521, U.S.A.}\\
\vspace*{1.5cm}
{\bf ABSTRACT} \\ \end{center}
\vspace*{5mm}
\noindent
The correlation between the two $\tau$-polarisations is predicted to be 
opposite $(+-/-+)$ for the $H/A \rightarrow \tau^+\tau^-$ signal, while it 
corresponds to the same sign combinations
for the Drell-Yan $(++/--)$ and $t \bar t$ $(--)$ 
backgrounds.  We show that this correlation can serve as a distinctive test 
to confirm the presence of the MSSM Higgs bosons $H$ and $A$ in their hadronic
$\tau^+\tau^-$ decay channels at the LHC.
\vfill\eject

\setcounter{page}{1}
\pagestyle{plain}

\noindent
The Minimal Supersymmetric Standard Model (MSSM) contains two Higgs doublets
\cite{Guide}.
The ratio of their vacuum expectation values is denoted by tan$\beta$,
which is required to be larger than 2 by the LEP data \cite{aa,bb}.
The two complex scalar doublets correspond to eight independent
fields, three of which are absorbed by the $W^\pm$ and $Z$ bosons. Thus we 
are left with five
physical states: two neutral scalars $h$ and $H$, a neutral pseudoscalar 
$A$, along
with a pair of charged Higgs bosons $H^\pm$. For large $M_A$ values 
($\gg M_Z$), the two heavy
neutral Higgs bosons, $H$ and $A$, have similar masses and couplings. 
On the one hand, the $H/A$ boson
couplings to the top quark are suppressed by a factor of cot$\beta$ relative
to that of the Standard Model
(SM) Higgs boson, while they both decouple from the $W^\pm$ and $Z$ bosons. 
On the other
hand, their couplings to the bottom quark are enhanced by a factor of 
tan$\beta$
relative to that of the SM Higgs boson. Thus one expects a large $H/A$ 
production
signal at the Large Hadron Collider (LHC) over the large tan$\beta$ region 
via the $b\bar b\to H/A$ `fusion' mechanism. 
Moreover the $H/A$ couplings to the $\tau$ lepton are
also enhanced by a factor of tan$\beta$ over this region.

Therefore the $H/A \rightarrow \tau^+\tau^-$ decay channel 
provides the most promising signature for the two heavier
neutral Higgs bosons of the MSSM at the LHC.  
Simulation studies have shown this 
decay channel to be viable over a large part of the parameter space
of the model  \cite{aa,bb}.  Moreover, a systematic study of this process in 
both the leptonic and hadronic decay channels of the $\tau$'s 
 \cite{cc,dd} shows that the latter has two 
advantages, i.e., a larger branching ratio (BR) 
and a larger visible fraction of
the $\tau$-energy/momentum.  This implies that one can reconstruct the 
$\tau^+\tau^-$ invariant mass most precisely via
\begin{equation}\label{one}
\tau^+\tau^- \rightarrow h^+h^-\nu\nu
\end{equation}
(hereafter, $h^\pm$ represents
a generic charged hadron)  \cite{cc}.  This channel 
is particularly useful for relatively large values of the two MSSM parameters
defining the Higgs sector of the MSSM at tree level, i.e., 
$M_A \gappeq 200~$GeV and $\tan \beta \gappeq 15$. This area is precisely
where the two Higgs states $H$ and $A$ are degenerate 
in their masses and couplings.

In this note, we shall analyse a distinctive feature of the 
$H/A\to \tau^+\tau^-$ signal
that has not received much attention so far, i.e., the correlation between 
the $\tau$-polarisations.  It simply follows from angular momentum 
conservation that the $\tau^+\tau^-$ pairs emerging from  the $H/A$ 
decays have opposite polarisations, i.e.,
\beq\label{two}
H/A \rightarrow \tau^+\tau^- X,~
P_{\tau^+}(P_{\tau^-}) = \pm 1(\mp 1).
\eeq
In contrast, the main irreducible backgrounds from Drell-Yan (DY) 
and top-antitop
processes predict the same
polarisation for the $\tau^+\tau^-$ pair, i.e.,
\beq
\gamma^*/Z \rightarrow \tau^+\tau^-~X,~
P_{\tau^+}(P_{\tau^-}) = \pm 1(\pm 1),
\label{three}
\eeq
\beq
t \bar t \rightarrow b\bar b W^+W^- \rightarrow \tau^+\tau^-~X,~
P_{\tau^+}(P_{\tau^-}) = -1(-1).
\label{four}
\eeq
The distinctive $\tau$-polarisations for the above signal and background 
processes were first discussed in Ref.~\cite{ee}, along with the analogous 
distinction between the charged Higgs signal and the $W^\pm$
background, i.e.,
\beq
H^\pm \rightarrow \tau^\pm\nu,~P_{\tau^\pm} = +1;~
W^\pm \rightarrow \tau^\pm\nu,~P_{\tau^\pm} = -1.
\label{five}
\eeq
The possibility of measuring the predicted $\tau$-polarisations via their 
hadronic decays was also discussed there.  Meanwhile, quantitative 
Monte Carlo (MC) studies of the charged Higgs boson signal at the 
LHC have shown 
the polarisation distinction in (\ref{five}) to be a very powerful tool
in separating the $H^\pm$ and $W^\pm$ contributions in the hadronic 
$\tau$-decay channels  \cite{ff,gg}.  
To the best of our knowledge, however, there 
has been no such quantitative study for the neutral Higgs signal in 
(\ref{two}).  
Of course, one can see from Eqs.~(\ref{two})--(\ref{four}) that the 
polarisation distinction for the neutral Higgs signal is more subtle than the 
charged Higgs case, Eq.~(\ref{five}).  In particular, the average 
$\tau$-polarisation for the $H/A \rightarrow \tau^+\tau^-$ signal is 0. 
The same is also true for the DY background in (\ref{three}), since 
the vector coupling of $\gamma$ and the dominantly axial coupling of $Z$ imply
that the polarisation combinations $++$ and $--$ occur with almost equal 
probability.  Nonetheless, we shall see below that the correlation 
between the $\tau$-polarisations can be effectively used to distinguish 
the signal from the background in the hadronic $\tau^+\tau^-$ channel.

Our results are based on a parton level MC simulation of the $H/A$ signal 
in (\ref{two}) as well as the $\gamma^*,Z$ and $t \bar t$ background 
processes in (\ref{three}) and (\ref{four}).  The dominant signal process in 
the large $\tan\beta~(\gappeq 15)$ region of interest can be emulated
via associate production with $b$-quarks \cite{fusion}:
\beq
gg, q \bar q \rightarrow b\bar b H/A.
\label{six}
\eeq
The cross sections for process (\ref{six})
have been calculated by several groups \cite{bbHiggs}. While these references
do not give compact formulae, the latter are very similar to those for
$gg, q \bar q \rightarrow t \bar b H^-$, described in detail in 
Ref.~\cite{hh}, which have been used again here, after rearranging the 
Higgs-to-fermion couplings appropriately. The formulae for the $H/A$ decay 
rates are taken from Ref.~\cite{ii} (see also \cite{others}).  
The DY and top-antitop background processes are 
computed using MadGraph \cite{jj}.  The phase space integrations have been done
numerically with the help of VEGAS  \cite{kk}.  We have used the parton 
distribution functions MRS99(COR01) \cite{MRS99}, evaluated at the scale 
$\mu = M_H(M_A)$ for the signal and $\mu = \sqrt{{\hat s}} /2 (2m_t)$ for the 
$\gamma^*,Z(t \bar t)$ backgrounds.  
(The results for these two processes
are however insensitive to the precise value of these scales.)
The `running' expression of the 
$b$-quark mass was used for the Yukawa coupling of the signal process.

To emulate detector resolution, we have applied Gaussian smearing on the 
transverse momentum of each parton jet (including the $\tau$-jets) with
\beq
(\sigma(p_T)/p_T)^2 = (0.6/\sqrt{p_T})^2 + (0.04)^2
\label{seven}
\eeq
in GeV units.  The missing transverse momentum, $p^{\rm {miss}}_T$, 
is reconstructed
 from the vector sum of the jet $p_T$'s after resolution smearing.  Following 
Ref.~\cite{cc}, we have imposed the following selection cuts on the two
$\tau$-jets and the accompanying missing transverse momentum:
\begin{eqnarray}\nonumber
\Delta R({\rm{jet}}-{\rm {jet}})>0.4,~&&~
p_T(\tau{\rm -jets}) > 60~{\rm GeV},~~~~~~~|\eta(\tau{\rm-jets})| < 2.5, \\ 
~~\Delta\phi(\tau{\rm-jets}) < 175^o,~&&~
 p^{\rm miss}_T > 40~{\rm GeV}.
\label{eight}
\end{eqnarray}
As in Ref.~\cite{cc}, we shall concentrate on the 1-prong hadronic decay 
channels of $\tau$'s, which are best suited for distinguishing $\tau$-jets 
from the QCD background.  With the above cuts, the QCD background has been 
estimated to be relatively small compared to the DY and top-antitop 
ones \cite{cc}.  While the DY background could be suppressed by 
$b$-tagging, it would also reduce the signal size significantly.  
Therefore, we do not require any $b$-tagging, just like Ref.~\cite{cc}.

The 1-prong hadronic decay channel of $\tau$'s 
accounts for 80\% of its hadronic
 decay width and 50\% of the total width.  The main contributors to the 1-prong 
hadronic decay are
\beq
\tau^\pm \rightarrow \pi^\pm\nu(12\%),~~\rho^\pm\nu(26\%),~~a^\pm_1\nu(8\%),
\label{nine}
\eeq
where the BRs for $\pi$ and $\rho$ include the small $K$ and $K^*$ 
contributions, respectively, which have identical polarisation effects 
\cite{ll}.  
The centre-of-mass  (CM) angular distribution of $\tau$-decays 
into $\pi$ or a vector meson $v(= \rho, a_1)$ is simply given in terms of its 
polarisation as
\beq
\frac{1}{\Gamma_\pi}\frac{d \Gamma_\pi}{d \cos \theta} = \frac{1}{2} 
(1 + P_\tau \cos\theta),
\label{ten}
\eeq
\beq
\frac{1}{\Gamma_v}~\frac{d\Gamma_{v~L,T}}{d\cos\theta} = \frac{\frac{1}{2} 
m^2_\tau,m^2_v}{m^2_\tau + 2m^2_v} (1 \pm P_\tau \cos\theta),
\label{eleven}
\eeq
where $L,T$ denote the longitudinal and transverse polarisation states of the 
vector meson.  The fraction $x$ of $\tau$-momentum in the laboratory frame
carried by its decay 
hadron is related to the polar angle $\theta$ via
\beq
x = \frac{1}{2} 
(1 + \cos\theta) + \frac{m^2_{\pi,v}}{2m^2_\tau}~(1 - \cos\theta),
\label{twelve}
\eeq
in the collinear approximation.  Thus the visible momentum of $\tau$, 
which defines the momentum of the $\tau$-jet, is
\beq
p_{\tau{\rm -jet}} \equiv p_{\rm hadron} = xp_\tau.
\label{thirteen}
\eeq
We see from Eqs.~(\ref{ten})--(\ref{thirteen}) that $P_\tau = +1$ gives a 
harder $\tau$-jet than $P_\tau = -1$.

We can also see from Eqs.~(\ref{ten})--(\ref{thirteen}) that the hard 
$\tau$-jet is dominated by the $\pi$ and the longitudinal vector meson 
$(\rho_L,a_{1L})$ contributions for $P_\tau = +1$, while it is dominated by 
the transverse $\rho$ and $a_1$ $(\rho_T,a_{1T})$ contributions for 
$P_\tau = -1$.  The two sets can be distinguished by exploiting the fact 
that the transverse $\rho$ and $a_1$ decays favour  even sharing of momentum 
between the decay pions, while the longitudinal $\rho$ and $a_1$ decays 
favour uneven sharing, where the charged pion carries very little or most of 
the meson momentum.  One can measure the fraction of the visible $\tau$-jet 
momentum, carried by the charged pion,
\beq
R = p_{\pi^\pm}/p_{\tau{\rm -jet}},
\label{fourteen}
\eeq
by combining the charged prong momentum measurement in the tracker with the 
calorimetric energy deposit of the $\tau$-jet.  Then the hard $\tau$-jet is 
predicted to show peaks at $R \simeq 0$ and 1 for $P_\tau = +1$ and at 
$R \simeq 0.4$ for $P_\tau = -1$~\cite{ee,ff}.  Note that the 
$\tau^\pm \rightarrow \pi^\pm\nu$ contribution adds substantially to the 
$R \simeq 1$ peak for $P_\tau = +1$, while it is suppressed for $P_\tau = -1$.
One can easily derive the above results quantitatively for $\rho_{L,T}$ decays;
but one has to assume a dynamical model for the $a_1$ decay for a quantitative 
result.

We shall adopt the model of Ref.~\cite{mm}, based on a conserved axial-vector 
current approximation, which describes the $a_1 \rightarrow 3\pi$ decay data 
very well.  A detailed account of the $\rho$ and $a_1$ decay formalisms 
including finite width effects can be found in Refs.~\cite{ee,ff}.  
A simple FORTRAN code for 1-prong hadronic decays of polarised $\tau$-leptons 
based on these formalisms can be obtained from one of the authors.  

Before discussing the results, let us briefly describe the reconstruction of 
the $\tau^+\tau^-$ invariant mass.  The $\Delta\phi$ cut of Eq.~(\ref{eight}) 
ensures that the transverse momenta of the two $\tau$-jets, $p_{T1}$ and 
$p_{T2}$, are not in the back-to-back configuration.  In this case, one can 
resolve the $p^{\rm miss}_T$ along their directions and add it to $p_{T1}$ 
and $p_{T2}$.  Scaling up the respective sums by the ratios $p_1/p_{T1}$ and 
$p_2/p_{T2}$ gives the reconstructed momenta of the $\tau^+\tau^-$ pair.  The 
resulting $\tau^+\tau^-$ invariant mass represents the $H/A$ mass for the signal 
process (\ref{two}) and the $\gamma^*/Z$ mass for the DY background 
(\ref{three}); but it does not represent any physical quantity for 
the $t \bar t$ background (\ref{four}), since the latter
contains additional sources of missing momentum.

The production cross sections of the $H$ and $A$ bosons are practically 
identical in the large $\tan\beta$ region of our interest.  As an illustration,
we have calculated the signal cross sections for
\beq
\tan\beta = 15~~~{\rm and}~~~M_A = M_H = 200(300)~{\rm GeV},
\label{fifteen}
\eeq
which is at the edge of the discovery limit for this channel \cite{cc,gg}.

Fig.~1 shows the signal and background cross sections against the 
reconstructed $\tau^+\tau^-$ invariant mass.  We see that a $M_{\tau\tau}$ cut of 
150--300(200--400)~GeV will suppress the DY and top-antitop backgrounds without
any appreciable loss of the 200(300)~GeV $H/A$ signal.  Therefore, we shall 
impose these invariant mass cuts in the reminder of our analysis.  Let us 
first consider the $M_{H,A}$ = 200~GeV case.  The signal and background 
cross sections for this case are shown on the top row of Tab.~1.  They are 
seen to be of similar size.  With an expected luminosity of 300~fb$^{-1}$ 
at the LHC, they correspond to about 1000 signal events over a background of 
roughly similar size.  This constitutes a comfortably large
$S/\sqrt B$ ratio.

\begin{table}[h]
\begin{center}
\begin{tabular}{|c||c|c|c|c|}
\hline
$M_{H,A}$ (GeV) & $b \bar b A$ & $b \bar b H$ & $\gamma^*, Z$ & $ t \bar t$ \\
\hline
200 & 1.58 & 1.57 & 3.03 & 0.73 \\
300 & 1.88 & 1.86 & 4.14 & 0.60\\
\hline
\multicolumn{5}{|c|}{$\tan\beta =15$}\\
\hline
\end{tabular}
\end{center}
\caption[]{Cross sections in femtobarns for Higgs signals and backgrounds in 
the $\tau^+\tau^- X$ channel, after the selection cuts of Eq.~(\ref{eight}).  
In addition, we have imposed a $M_{\tau\tau}$ cut of 150--300~GeV for 
$M_{H,A}$ = 200~GeV and 200--400~GeV for $M_{H,A}$ = 300~GeV.  We have also 
imposed a stronger $p_T$ cut of 100~GeV on the harder $\tau$-jet in the 
latter case.}
\end{table}

Fig.~2 shows the three-dimensional (3-D) plots of the 200~GeV $H/A$ signals 
along with the DY and $t \bar t$ backgrounds against $R_1, R_2$, the 
fractional $\tau$-jet momenta carried by the charged prongs.  The subscript 
1(2) refers to the harder(softer) of the two $\tau$-jets.  There are several 
points worth noting in this figure.  The $\tau$-identification will require a 
minimum hardness for the charged prongs, presumably $R_1$ and $R_2 > 0.2$.  
Thus the $R_{1,2} < 0.2$ contributions from $\rho_L$ and $a_{1L}$  will not be
relevant for the practical analysis.  However, the peaks at $R_{1,2} 
\simeq 0.4$ and $R_{1,2} \simeq 1$, coming from the $\rho_T$, $a_{1T}$ and 
$\rho_L,\pi$ respectively, can be used to distinguish between the 
$P_\tau = -1$ and +1 contributions.  In particular, the $t \bar t$ background,
corresponding to the polarisation combination $--$, is crowded around $R_{1,2} 
\simeq 0.4$.  In contrast, the signal is peaked at $R_1$ or $R_2 > 0.8$, as 
expected for the polarisation combination $+-$.  The same is seen to be true 
for the DY background, because the $p_T > 60$~GeV cut on the 
$\tau$-jets has suppressed the $--$ contribution with respect to the $++$ 
combination.  Thus requiring at least one of the two $\tau$-jets to contain a 
very hard charged prong, carrying $> 80\%$ of its visible momentum, will 
suppress the $t \bar t$ background effectively without any significant loss 
to the signal or the DY background.  Moreover, we expect this requirement to 
suppress the QCD background to an even greater extent than the $t \bar t$ 
background, because its $R$ distributions are softer than those
of the latter \cite{cc}.
Therefore, we strongly advocate an asymmetric cut, requiring at least one of 
the $R_1$ and $R_2$ to be $>0.8$, for effective suppression of the $t \bar t$ 
and QCD backgrounds (including $W$ + jets).  Then the signal can be 
distinguished from the remaining DY background by looking at the asymmetric 
region: $R_1 > 0.8$ and $R_2 = 0.2-0.8$ and vice versa.  The signal(DY 
background) has a peak(dip) in this region as expected from the 
polarisation correlation $+-$($++$).  
We have estimated 55\% of the signal events 
to come from this asymmetric region as against 43\% for the DY background.  
Thus excluding the symmetric region will improve the 
$S/B$ ratio, but not the $S/\sqrt B$.  On the other hand, the distribution of 
events between the symmetric and asymmetric regions can be used to 
distinguish the signal from the DY background.  From an event sample of 
$\sim1000$ one expects 550 to populate the above asymmetric region for the signal ($+-$) as against 430 for the background ($++$), which constitutes a 
$\gappeq 5\sigma$ excess.

For higher $H/A$ masses it may be more advantageous to use asymmetric $p_T$ 
cuts on the two $\tau$-jets.  Fig.~3 shows the 300~GeV $H/A$ signal along 
with the corresponding DY and $t \bar t$ backgrounds against the $p_T$ of the 
harder $\tau$-jet.  It shows that increasing the $p_{T1}$ cut to 100~GeV 
(while keeping the $p_{T2}$ cut at 60~GeV) will suppress the $t \bar t$ 
background to a much higher degree than the signal, as expected from their 
polarisation combinations, $--$ and $-+$, respectively.  The DY background is 
also suppressed to a somewhat larger extent than the signal, though the 
difference in rather modest (7--8\%) in this case.  The $p_{T2} > 
60$~GeV cut is already suppressing the 
$--$ contribution with respect to the $++$ 
combination even for $M_{\tau\tau} \simeq 300$~GeV.  We expect this 
difference to increase for higher $H/A$ masses.  We also expect the higher 
$p_{T1}$ cut to be very effective in suppressing the QCD background
 (again, including $W$+jets).  
Tab.~1 shows the 300~GeV $H/A$ signal cross sections 
along with the DY and $t \bar t$ backgrounds for this asymmetric cut (i.e.,
$p_{T1} > 100$~GeV, $p_{T2} > 60$~GeV).  Their sizes are similar to the 
200~GeV signal and background cross sections with symmetric $p_T$ cuts.  
We expect the asymmetric $p_T$ cuts to be advantageous in probing 
the region $M_{A/H} \gappeq 300$~GeV.

Fig.~4 shows the 3-D plots of the 300~GeV $H/A$ signal along with the DY 
and $t \bar t$ backgrounds against $R_1$ and $R_2$ for the asymmetric $p_T$ 
cuts mentioned above.  Most of the observations made for Fig.~2 apply to this 
case as well.  In particular, one sees a peak(dip) for the 
signal(DY background) at $R_1 > 0.8$ and $R_2 = 0.2-0.8$, as expected from the 
polarisation correlation $+-$($++$).  Note, however, that 
there is no clear peak 
for the signal in the complementary region, $R_2 > 0.8$ and $R_1 = 0.2-0.8$. 
This is because in this case the $p_{T2} > 60$~GeV cut is not hard 
enough to suppress the pion peak (at $R_2 = 1$) for $P_{\tau_2} = -1$.  
Consequently, the accompanying $\tau_1$ has a significant component of 
$P_{\tau_1} = +1$.  Thus for higher $H/A$ masses one should look for the 
signal peak in the $R_1 > 0.8$ and $R_2 = 0.2-0.8$ region only.  Nonetheless
the difference is very significant in this case. About 30\% of the 300 GeV
$H/A$ signal events populate this asymmetric region against 20\% of the
DY background events. This constitutes a 7$\sigma$ excess for an event
sample of $\sim1000$, which can be used as a confirmatory test of the $H/A$ signal.

Finally, notice that the percentages of the signal and 
 background events populating the asymmetric region are independent of
the normalisation uncertainties of the various
processes, since they depend entirely on the
polarisations of the latter. The former mainly stem from higher order 
corrections
due to QCD. While these are relatively well under control for the two 
backgrounds
considered here, they can be very large for the signal (see \cite{scale}
for an up-to-date discussion).

In summary, we have found that the correlation between the $\tau$-polarisations
are likely to play a very useful role in distinguishing the $H/A$ signal from 
the backgrounds in the hadronic $\tau^+\tau^-$ channel at LHC. In view of the
importance of this channel in the MSSM Higgs search programme at the LHC, it 
is imperative that this analysis is followed up with a more rigorous MC
simulation, incorporating jet hadronisation and detector acceptance effects.  
We have currently undertaken such a simulation study \cite{nn} using the 
HERWIG event generator \cite{oo,pp} interfaced with the TAUOLA package 
\cite{qq,rr} for polarised $\tau$-decays.

We gratefully acknowledge discussions with P. Aurenche, D. Denegri., R. Kinnunen
and A. Nikitenko.

\clearpage
\begin{figure}[!t]
\begin{center}
~{\epsfig{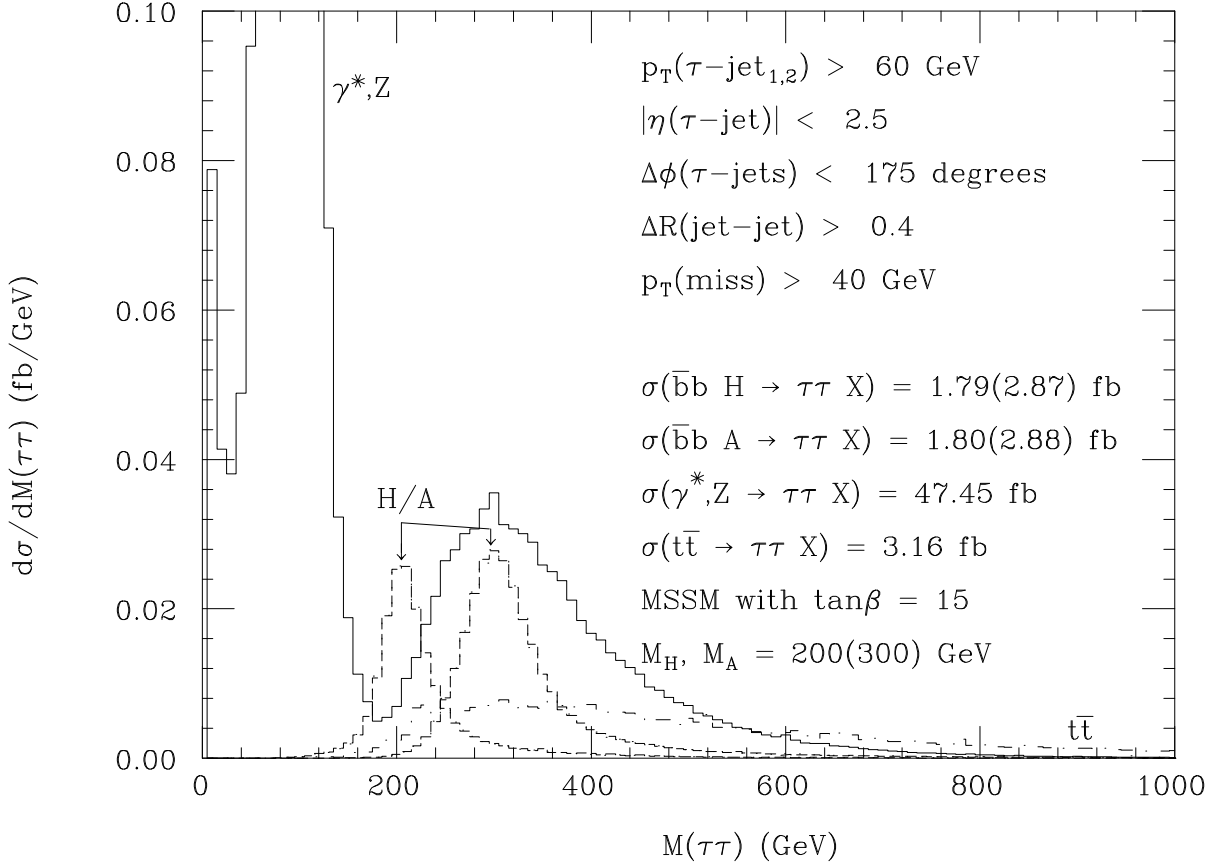}}
\end{center}
\begin{center}
\caption{Singly differential distributions in the reconstructed
$\tau^+\tau^-$ invariant mass in 1-prong hadronic decays, for the Higgs
signals (dashed and dotted lines for $H$ and $A$, respectively, 
coinciding visually) 
and backgrounds (solid and dot-dashed lines for DY and top-antitop,
respectively) 
in the $\tau^+\tau^-X$ channel, after the
following selection cuts (at parton level):
$p_T(\tau-{\mathrm{jet}}_{1[2]}) >  60[60]$    GeV (1[2] refers
to the most[least] energetic),
$|\eta(\tau-{\mathrm{jet}})| <  2.5$,
$\Delta\Phi(\tau-{\mathrm{jets}}) <  175^o$,
$\Delta R({\mathrm{jet}}{\mathrm{-jet}}) >  0.4$ and
$p_T^{\mathrm{miss}} >  40$    GeV.
Normalisations are to the total cross sections  
at $\sqrt s=14$ TeV (for $M_A=M_H=200(300)$ GeV and $\tan\beta=15$,
in the case of the signal). 
Bins are 10 GeV wide.}
\label{fig:mass}
\end{center}
\end{figure}

\clearpage

\begin{figure}[htb]
\begin{center}
\centerline{\framebox{$d^2\sigma/dR_1/dR_2/\sigma$}}
\vskip1.0cm
\begin{minipage}[b]{.5\linewidth}
\centerline{$b\bar b A\to \tau^+\tau^- X$ ($M_A=200$ GeV)}
\vspace{-0.75truecm}\centering\epsfig{file=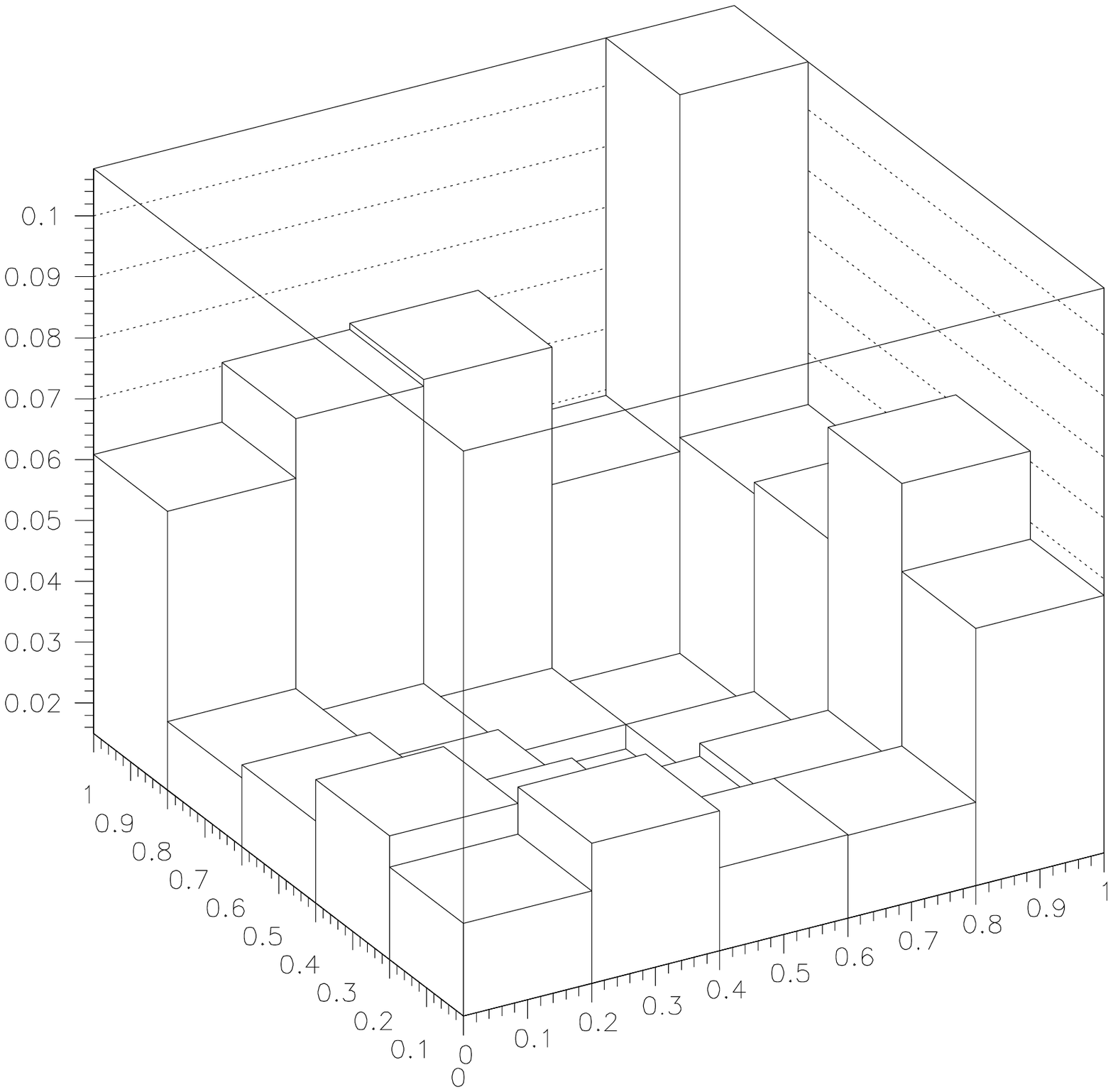,angle=0,height=8cm,width=\linewidth}
\vskip-2.0cm
\centerline{
$R_1$
\qquad\qquad\qquad\qquad\qquad\qquad\quad
$R_2$}
\end{minipage}\hfil
\begin{minipage}[b]{.5\linewidth}
\centerline{$b\bar b H\to 
\tau^+\tau^- X$ ($M_H=200$ GeV)}
\vspace{-0.75truecm}\centering\epsfig{file=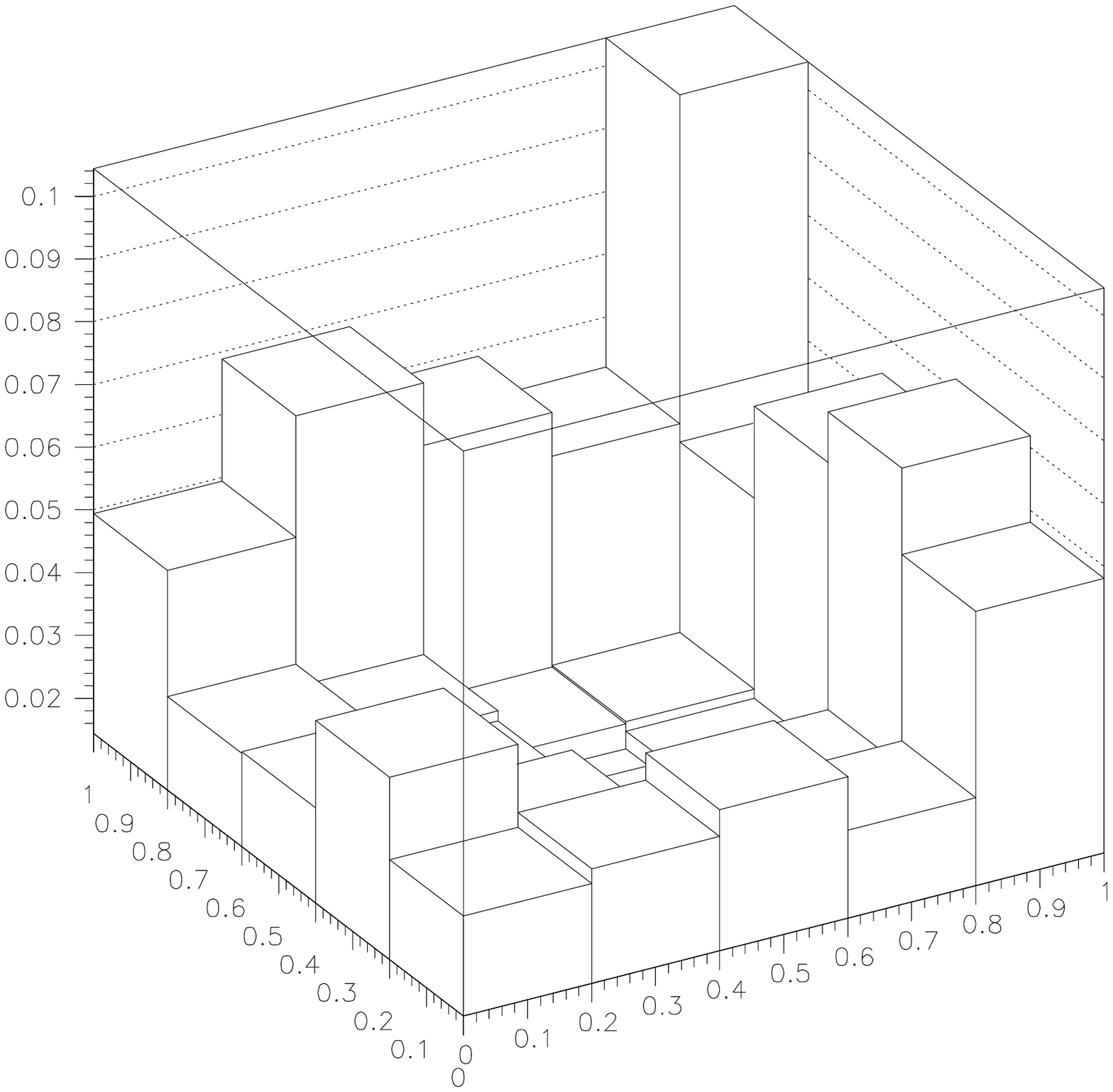,angle=0,height=8cm,width=\linewidth}
\vskip-2.0cm
\centerline{
$R_1$
\qquad\qquad\qquad\qquad\qquad\qquad\quad
$R_2$}
\end{minipage}\hfil
\centerline{}
\begin{minipage}[b]{.5\linewidth}
\vspace{+1.0truecm}\centerline{$t\bar t\to \tau^+\tau^- X$}
\vspace{-0.75truecm}\centering\epsfig{file=    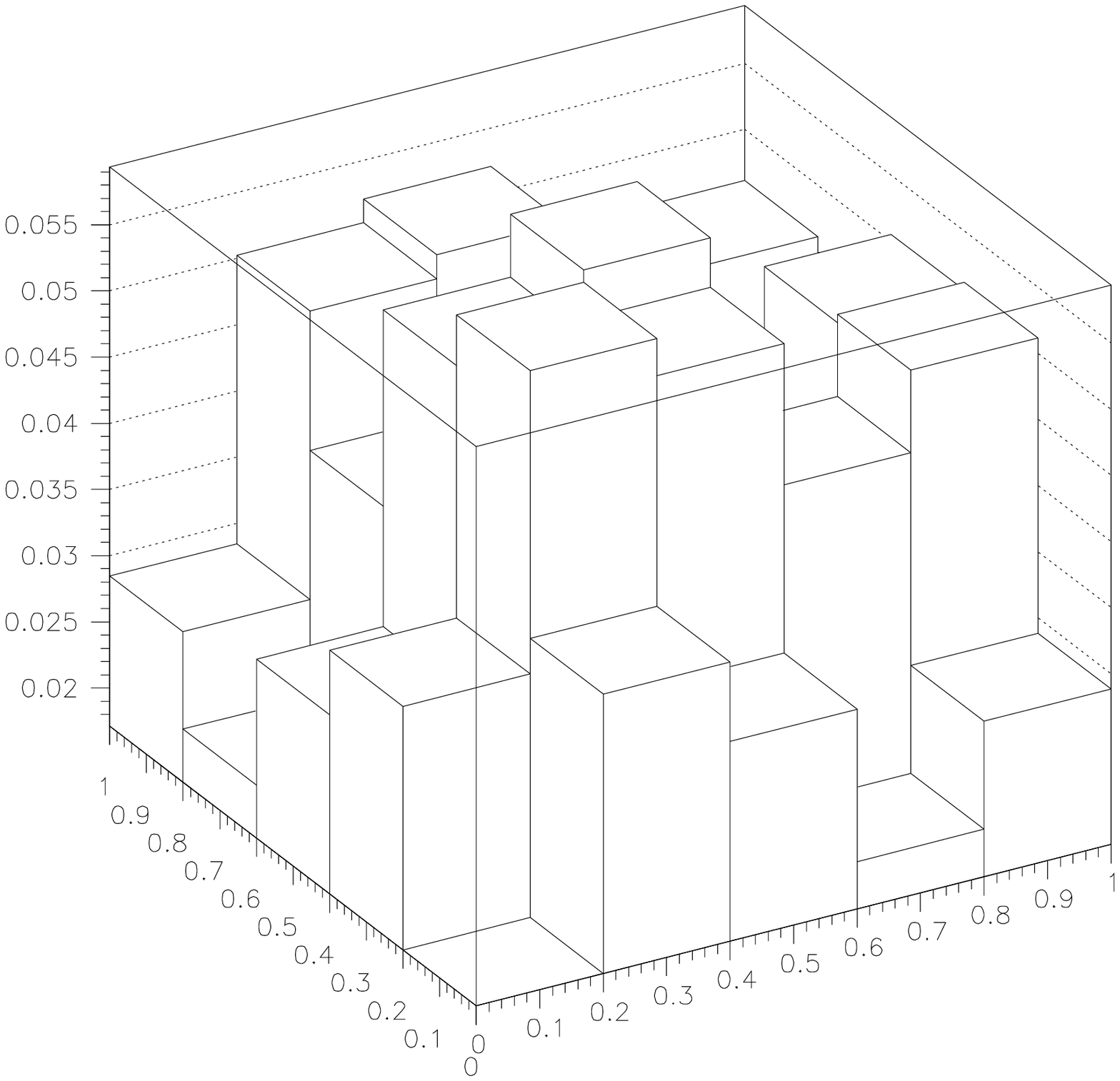,angle=0,height=8cm,width=\linewidth}
\vskip-2.0cm
\centerline{
$R_1$
\qquad\qquad\qquad\qquad\qquad\qquad\quad
$R_2$}
\end{minipage}\hfil
\begin{minipage}[b]{.5\linewidth}
\vspace{+1.0truecm}\centerline{$\gamma^*,Z\to 
\tau^+\tau^- X$}
\vspace{-0.75truecm}\centering\epsfig{file=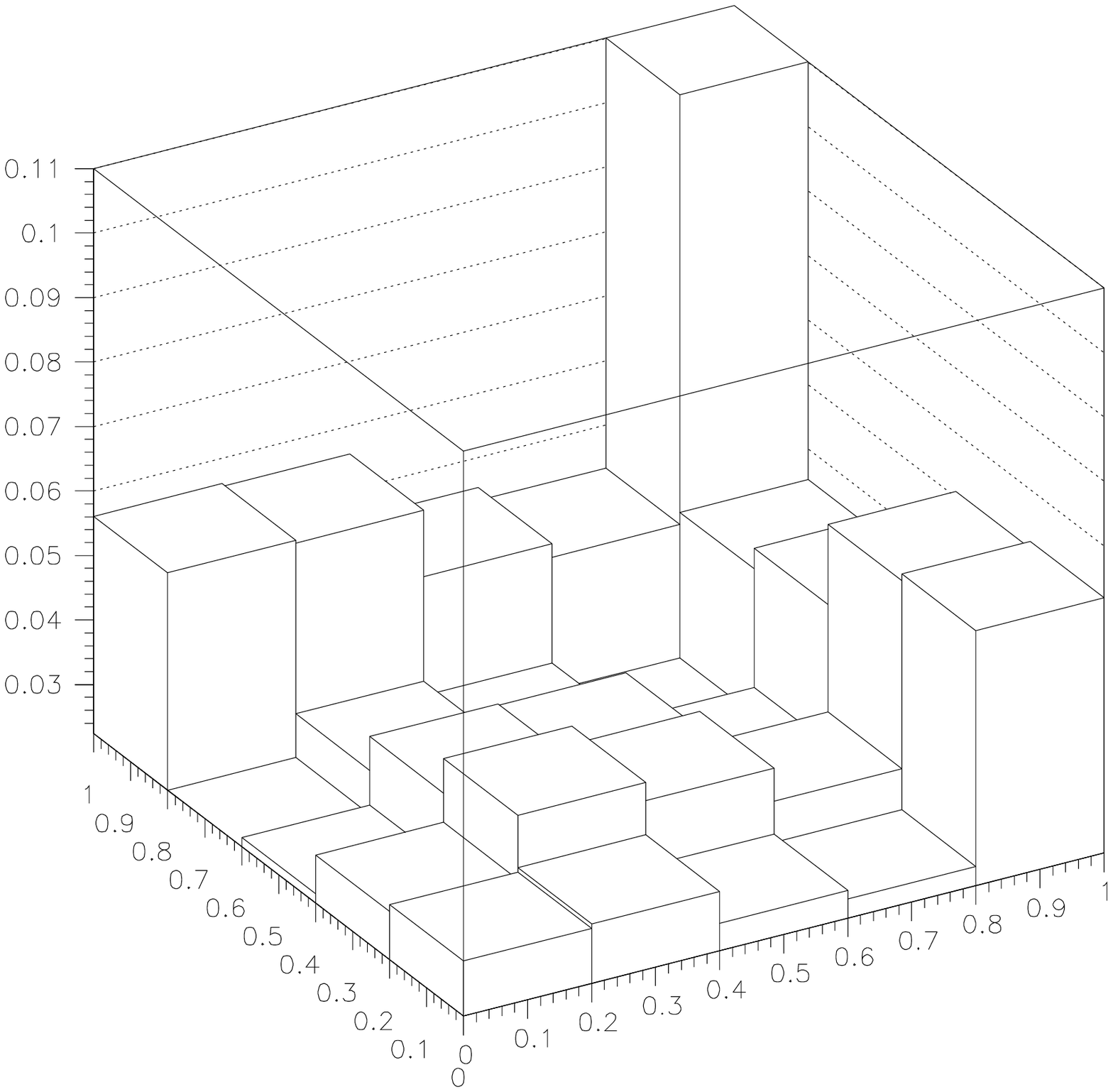,angle=0,height=8cm,width=\linewidth}
\vskip-2.0cm
\centerline{
$R_1$
\qquad\qquad\qquad\qquad\qquad\qquad\quad
$R_2$}
\end{minipage}\hfil
\centerline{}

\noindent
\caption{Doubly differential distributions in the energy fractions
$R_i = p^{\pi^\pm}_i / p_i^{\tau-\mathrm{jet}}$, where 
$p^{\pi^\pm}_i$ is the charged pion momentum and $p_i^{\tau-\mathrm{jet}}$ 
that of the visible $\tau-$jet, i.e, the momentum carried 
away by the mesons $\pi^\pm,
\rho^\pm$ and $a_1^\pm$ in 1-prong decays, for 
Higgs signals and backgrounds in the $\tau^+\tau^-X$ channel
(with $p_1>p_2$), after the
following selection cuts (at parton level):
$p_T(\tau-{\mathrm{jet}}_{1[2]}) >  60[60]$    GeV (1[2] refers
to the most[least] energetic),
$|\eta(\tau-{\mathrm{jet}})| <  2.5$,
$\Delta\Phi(\tau-{\mathrm{jets}}) <  175^o$,
$\Delta R({\mathrm{jet}}{\mathrm{-jet}}) >  0.4$,
$p_T^{\mathrm{miss}} >  40$    GeV
and 150 GeV $<M_{\tau\tau}<300$ GeV.
Normalisations are to unity
at $\sqrt s=14$ TeV (for $M_A=M_H=200$ GeV
in the case of the signal). 
Bins are 0.2 units wide.}
\label{fig:measure1}
\end{center}
\end{figure}

\clearpage

\begin{figure}[!t]
\begin{center}
~{\epsfig{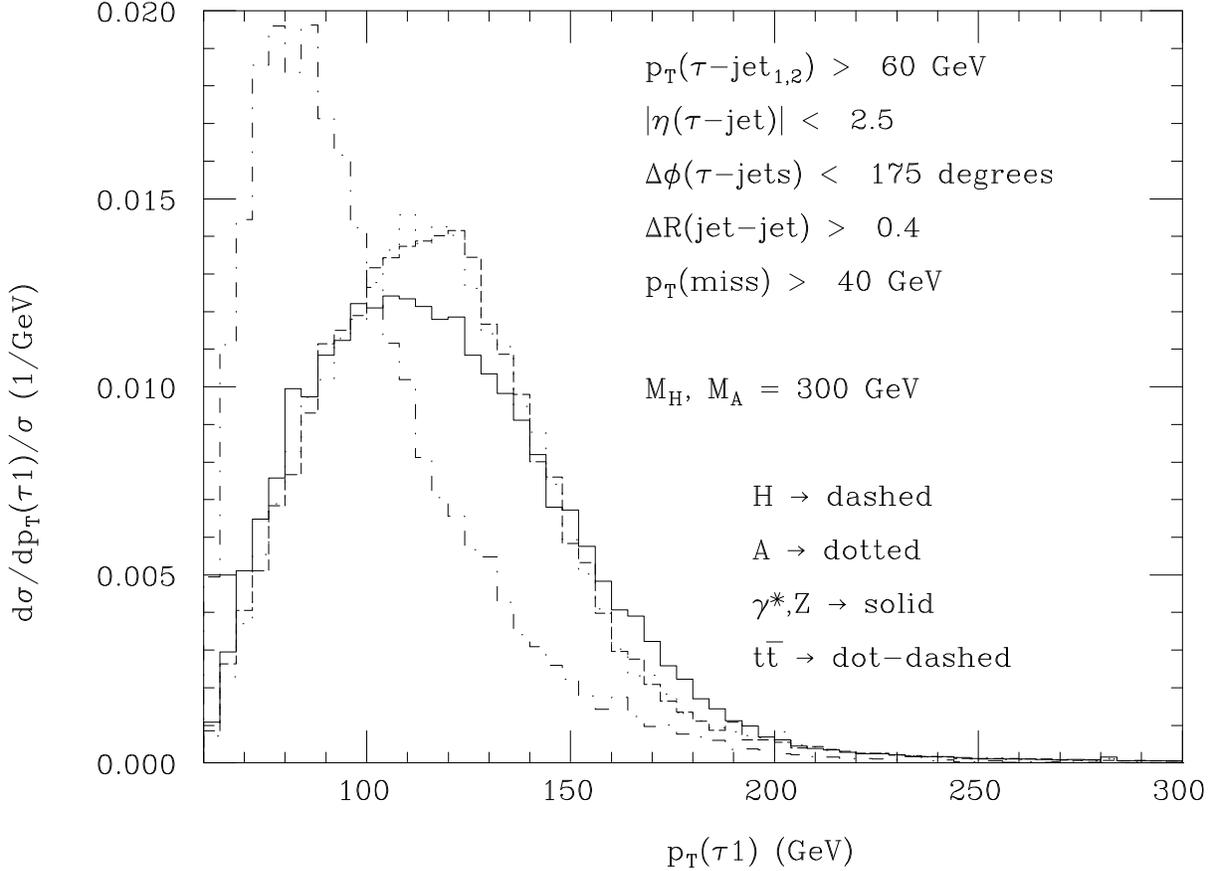}}
\end{center}
\begin{center}
\caption{Singly differential distributions in the transverse momentum
of the most energetic $\tau$-jet in 1-prong decays, for 
Higgs signals and backgrounds in the $\tau^+\tau^-X$ channel, after the
following selection cuts (at parton level):
$p_T(\tau-{\mathrm{jet}}_{1[2]}) >  60[60]$    GeV (1[2] refers
to the most[least] energetic),
$|\eta(\tau-{\mathrm{jet}})| <  2.5$,
$\Delta\Phi(\tau-{\mathrm{jets}}) <  175^o$,
$\Delta R({\mathrm{jet}}{\mathrm{-jet}}) >  0.4$ and
$p_T^{\mathrm{miss}} >  40$    GeV.
Normalisations are to unity  
at $\sqrt s=14$ TeV (for $M_A=M_H=300$ GeV
in the case of the signal). 
Bins are 4 GeV wide.}
\label{fig:pT}
\end{center}
\end{figure}

\clearpage

\begin{figure}[htb]
\begin{center}
\centerline{\framebox{$d^2\sigma/dR_1/dR_2/\sigma$}}
\vskip1.0cm
\begin{minipage}[b]{.5\linewidth}
\centerline{$b\bar b A\to \tau^+\tau^- X$ ($M_A=300$ GeV)}
\vspace{-0.75truecm}\centering\epsfig{file=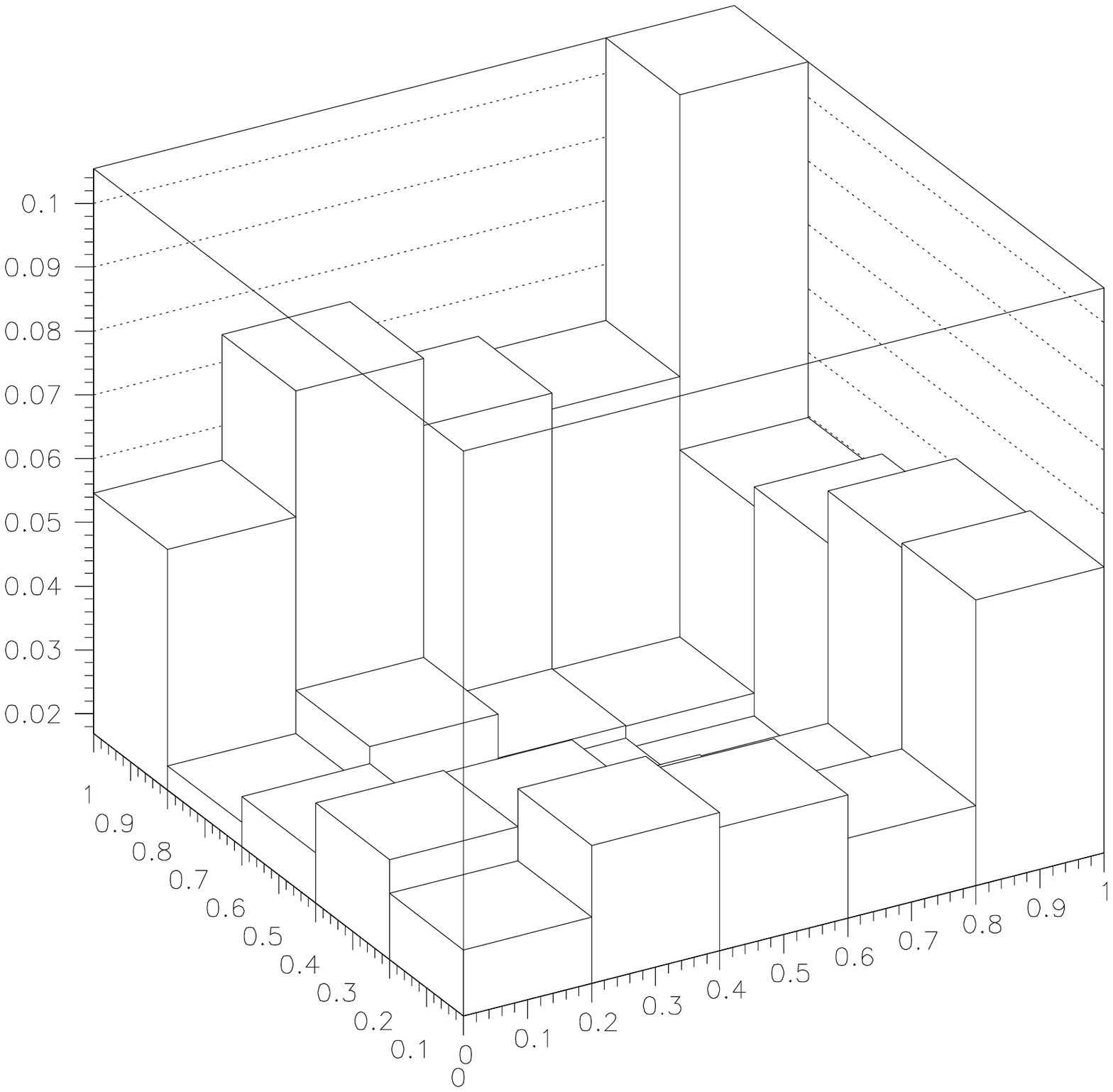,angle=0,height=8cm,width=\linewidth}
\vskip-2.0cm
\centerline{
$R_1$
\qquad\qquad\qquad\qquad\qquad\qquad\quad
$R_2$}
\end{minipage}\hfil
\begin{minipage}[b]{.5\linewidth}
\centerline{$b\bar b H\to 
\tau^+\tau^- X$ ($M_H=300$ GeV)}
\vspace{-0.75truecm}\centering\epsfig{file=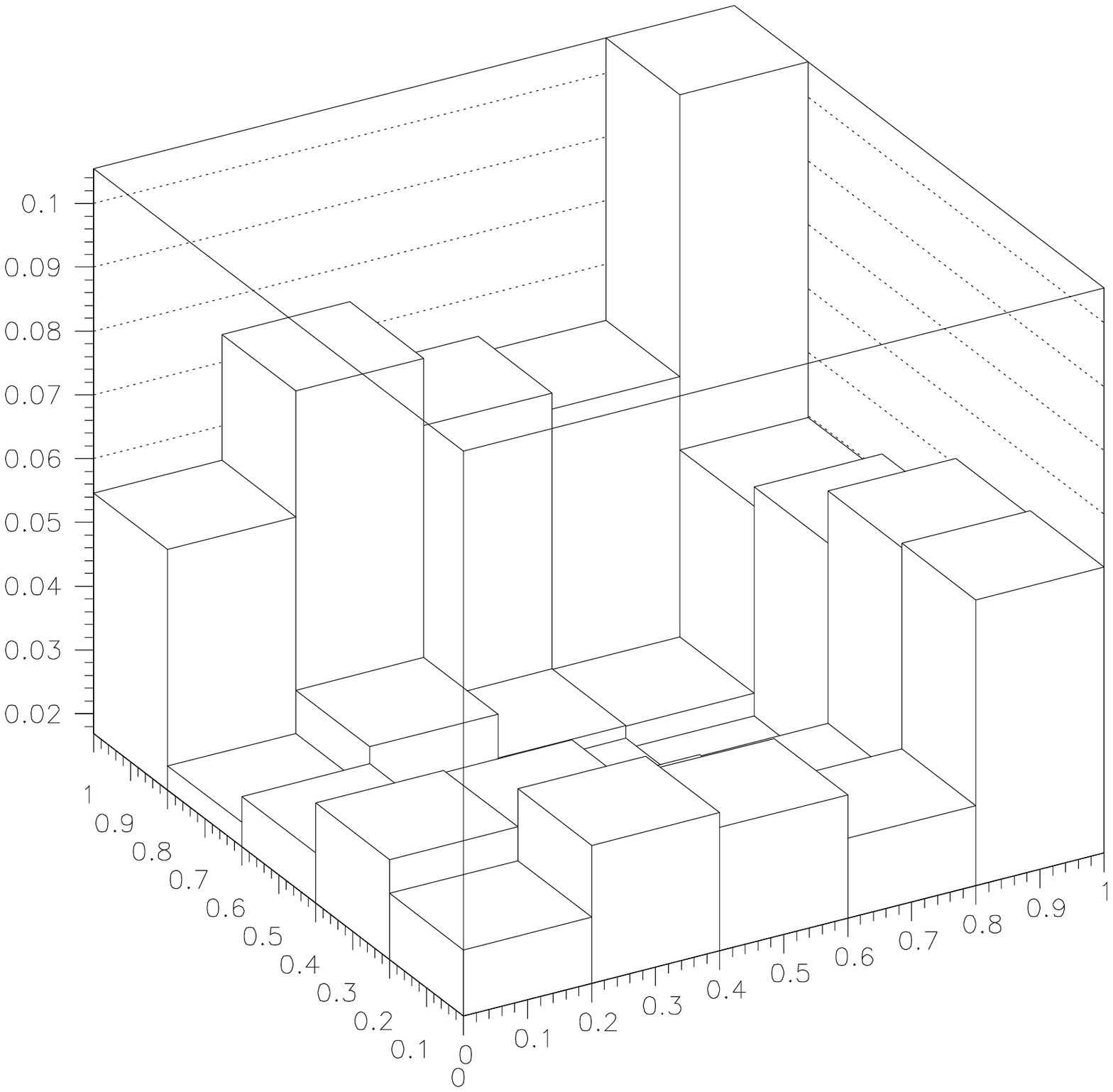,angle=0,height=8cm,width=\linewidth}
\vskip-2.0cm
\centerline{
$R_1$
\qquad\qquad\qquad\qquad\qquad\qquad\quad
$R_2$}
\end{minipage}\hfil
\centerline{}
\begin{minipage}[b]{.5\linewidth}
\vspace{+1.0truecm}\centerline{$t\bar t\to \tau^+\tau^- X$}
\vspace{-0.75truecm}\centering\epsfig{file=    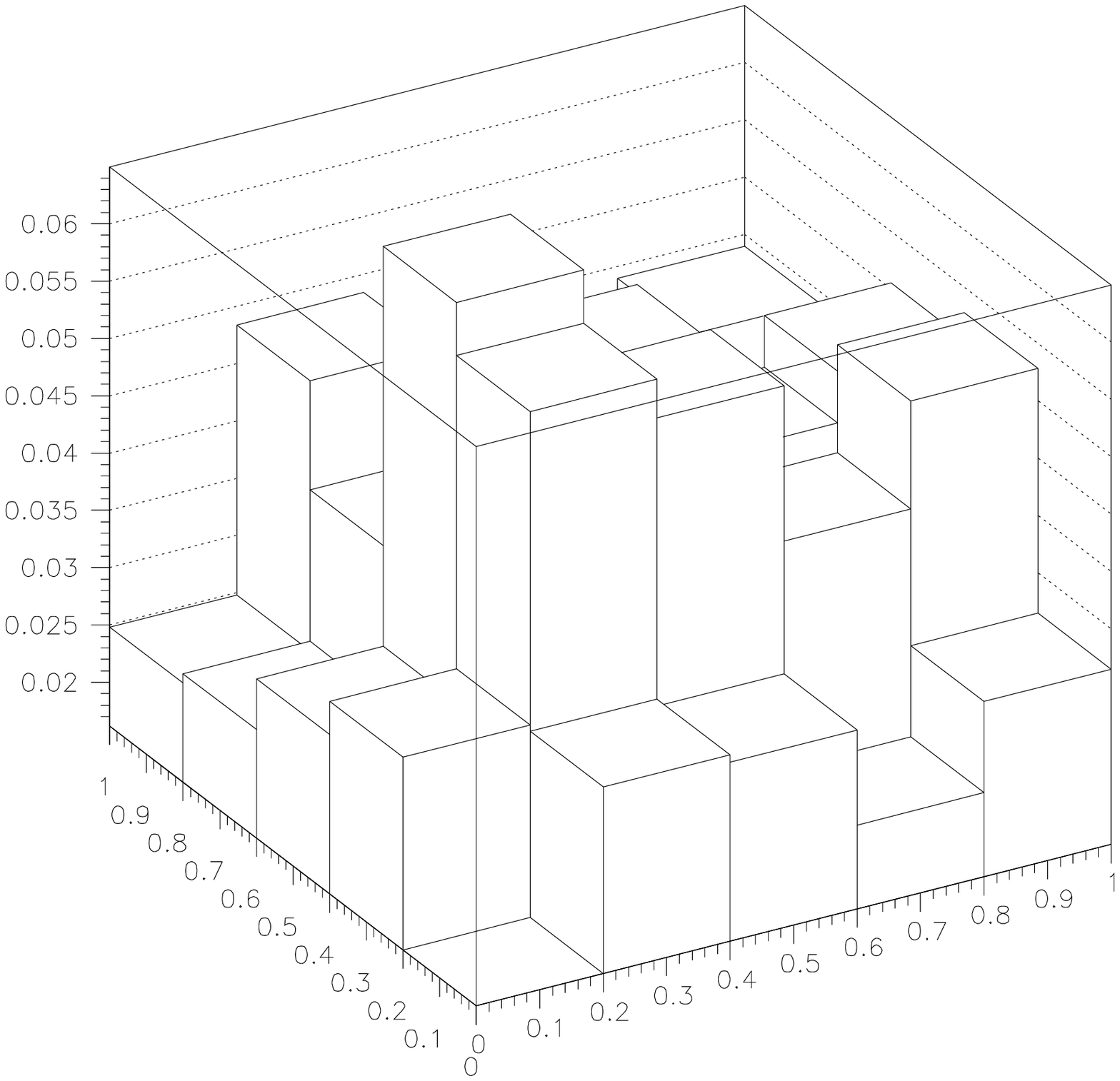,angle=0,height=8cm,width=\linewidth}
\vskip-2.0cm
\centerline{
$R_1$
\qquad\qquad\qquad\qquad\qquad\qquad\quad
$R_2$}
\end{minipage}\hfil
\begin{minipage}[b]{.5\linewidth}
\vspace{+1.0truecm}\centerline{$\gamma^*,Z\to 
\tau^+\tau^- X$}
\vspace{-0.75truecm}\centering\epsfig{file=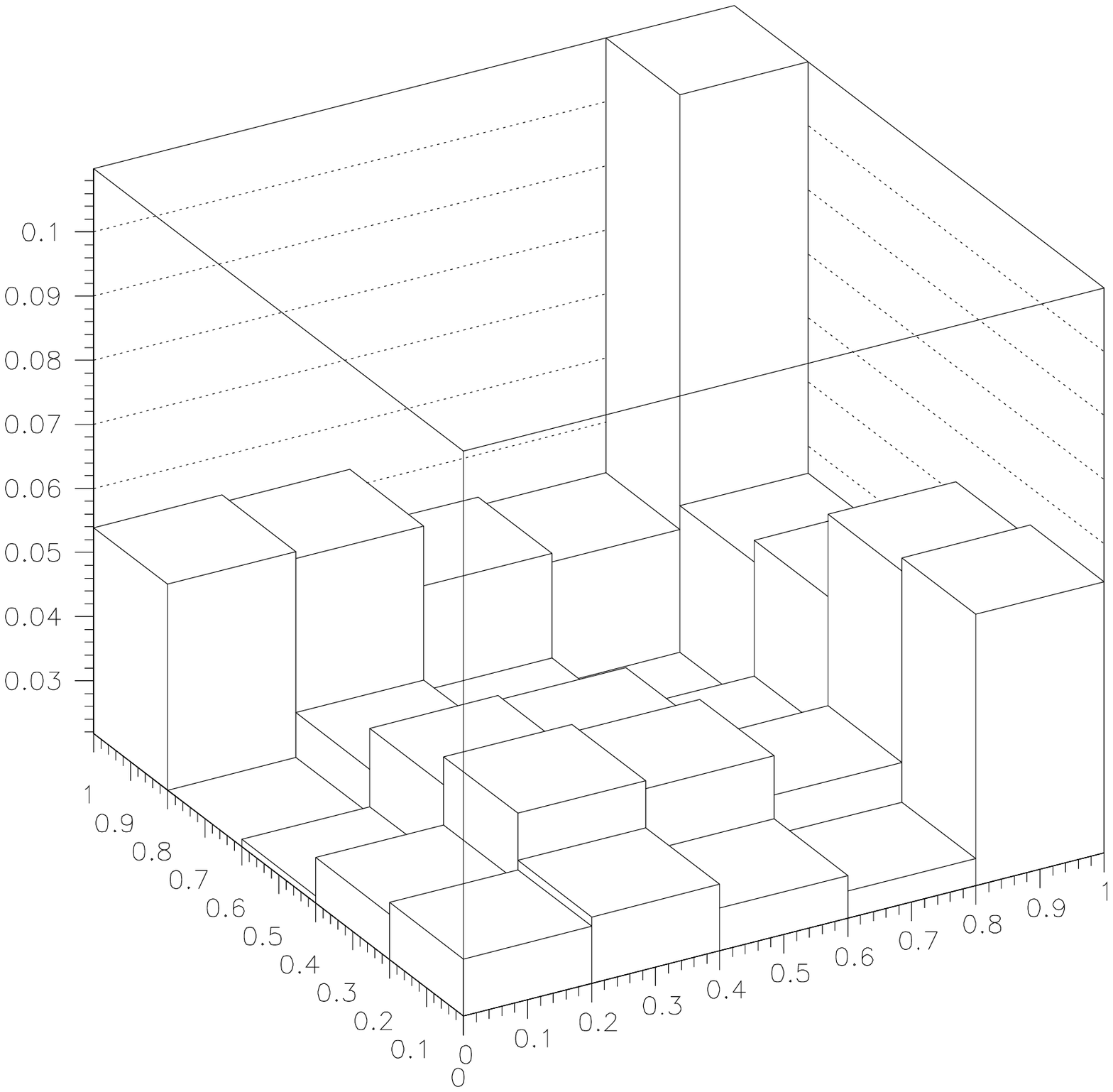,angle=0,height=8cm,width=\linewidth}
\vskip-2.0cm
\centerline{
$R_1$
\qquad\qquad\qquad\qquad\qquad\qquad\quad
$R_2$}
\end{minipage}\hfil
\centerline{}

\noindent
\caption{Doubly differential distributions in the energy fractions
$R_i = p^{\pi^\pm}_i / p_i^{\tau-\mathrm{jet}}$, where 
$p^{\pi^\pm}_i$ is the charged pion momentum and $p_i^{\tau-\mathrm{jet}}$ 
that of the visible $\tau-$jet, i.e, the momentum carried 
away by the mesons $\pi^\pm,
\rho^\pm$ and $a_1^\pm$ in 1-prong decays, for 
Higgs signals and backgrounds in the $\tau^+\tau^-X$ channel
(with $p_1>p_2$), after the
following selection cuts (at parton level):
$p_T(\tau-{\mathrm{jet}}_{1[2]}) >  100[60]$    GeV (1[2] refers
to the most[least] energetic),
$|\eta(\tau-{\mathrm{jet}})| <  2.5$,
$\Delta\Phi(\tau-{\mathrm{jets}}) <  175^o$,
$\Delta R({\mathrm{jet}}{\mathrm{-jet}}) >  0.4$,
$p_T^{\mathrm{miss}} >  40$    GeV
and 200 GeV $<M_{\tau\tau}<400$ GeV.
Normalisations are to unity  
at $\sqrt s=14$ TeV (for $M_A=M_H=300$ GeV 
in the case of the signal). 
Bins are 0.2 units wide.}
\label{fig:measure2}
\end{center}
\end{figure}

\end{document}